\documentclass[twocolumn]{aastex631}

\usepackage{multirow}
\graphicspath{{./}{figures/}}
\usepackage{amsmath}

\usepackage{CJK}

\newcommand\sref[1]{\hyperref[#1]{\S~\ref*{#1}}}
\newcommand\fref[1]{\hyperref[#1]{Fig.~\ref*{#1}}}
\newcommand\Eqref[1]{Eq.~(\hyperref[#1]{\ref*{#1}})}
\newcommand\eeqref[1]{Eq.~\hyperref[#1]{\ref*{#1}}}
\newcommand\tref[1]{\hyperref[#1]{Table~\ref*{#1}}}
\newcommand\aref[1]{\hyperref[#1]{Appendix~\ref*{#1}}}

\shorttitle{Simulating Bow Shocks of Runaway SMBH}
\shortauthors{Su et al.}

\begin{document}
\begin{CJK}{UTF8}{mj}

\title{Shocks and Trails in the Circumgalactic Medium Driven by Runaway Supermassive Black Holes}

\author[0000-0003-1598-0083]{Kung-Yi Su}
\affiliation{Department of Physics \& Astronomy and Center for Interdisciplinary Exploration and Research in Astrophysics (CIERA), Northwestern University, 1800 Sherman Ave, Evanston, IL 60201, USA}
\email{kungyisu@gmail.com}
\author[0000-0002-4900-6628]{Claude-Andr\'e Faucher-Gigu\`ere }
\affiliation{Department of Physics \& Astronomy and Center for Interdisciplinary Exploration and Research in Astrophysics (CIERA), Northwestern University, 1800 Sherman Ave, Evanston, IL 60201, USA}
\begin{abstract}
\label{abstract}
Runaway supermassive black holes (SMBHs) have long been predicted.
RBH-1, discovered with HST and confirmed by JWST, provides the strongest evidence yet for a runaway SMBH: a bow shock and linear trail in the circumgalactic medium (CGM) of a $z=0.96$ galaxy.
The bow shock is $\sim$kpc wide, the trail is $\sim 60$ kpc long, and the SMBH apparently moves at $\sim 1,000$ km s$^{-1}$.
Using controlled hydrodynamic simulations, we show that the SMBH's gravity alone cannot produce a significant bow shock, but a comoving wind source can sustain one similar to RBH-1's.
The wind must either be active or have shut off less than $\sim 3$ Myr ago.
The required recent wind momentum flux is $6\times10^{32}\,{\rm dyn}\,(n_{\rm CGM}/10^{-3}\,{\rm cm}^{-3})$, where $n_{\rm CGM}$ is the CGM density.
Assuming the wind reaches peak strength only during the last $\sim 6$ Myr, the minimum integrated wind mass is $3.5\times10^5\,M_\odot\,(n_{\rm CGM}/10^{-3}\,{\rm cm}^{-3})$.
A $2\times10^7\,M_\odot$ SMBH accreting at $1\%$ Eddington could power the wind.
Supernova feedback from a star cluster carried by the SMBH is an alternative, although the required mass and momentum fluxes $\gtrsim 60$ Myr after ejection are difficult to reconcile with a standard initial mass function.
For a hot ambient CGM at $\sim 10^6$ K, the quasi-linear trail constrains the 1D velocity dispersion to $\sigma_{\rm 1D}\lesssim 30$ km s$^{-1}$, above which it becomes strongly disturbed.
Runaway SMBH trails therefore offer a new probe of turbulence in the diffuse CGM, in regimes where its kinematics are otherwise inaccessible.

\end{abstract}

\section{Introduction}
\label{S:intro}
Runaway supermassive black holes (SMBHs) have long been theoretically predicted to arise from either three-body slingshot interactions or gravitational recoil following black hole mergers. Despite the scarcity of direct observational evidence, a promising runaway SMBH candidate was identified in HST observations by \cite{2023ApJ...946L..50V} and subsequently confirmed by follow-up JWST observations \citep{2026ApJ...998L..27V}. These observations reveal a shock front located $\sim 60$ kpc from the host galaxy in the circumgalactic medium (CGM), with a width of $\sim 1$--$2$ kpc, potentially attributed to a $\sim 2\times 10^7 M_\odot$ black hole with a runaway velocity of $\sim 1000$ km s$^{-1}$. A linear trail also extends from the shock front to the host galaxy.

A runaway SMBH moving at such a velocity could plausibly be produced by gravitational-wave recoil following an SMBH merger with the progenitor properties inferred by \cite{2026PhRvL.137c1404I}.

Several theoretical and simulation studies have attempted to explain the observed features. \cite{2026ApJ...998L..27V} attributed the shock front to a momentum-driven scenario induced by the black hole itself without explicit outflows. \cite{2023ApJ...954L...2C} searched the ASTRID simulation for analogues of such a system and found that an encounter between a compact galaxy hosting an SMBH and a star-forming galaxy may produce a similar linear feature in the trail.

Regarding the formation of a linear filament along the trail, \cite{2024MNRAS.527.5503O} investigated whether the gravitational perturbation of the CGM by a passing runaway SMBH can induce converging gas flows that subsequently cool and condense. They conclude that a relatively low initial CGM temperature, $T\lesssim2\times10^5\,{\rm K}$, and a sufficiently high density, $n_{\rm H}\gtrsim2\times10^{-5}(T/2\times10^5\,{\rm K})^{-1}\,{\rm cm}^{-3}$, are required.
Recently, \cite{2026MNRAS.550g1275K} further used a set of idealized hydrodynamic simulations to study how turbulent mixing and radiative cooling can reproduce the cold gas in the trail, with the bow shock itself created by an object put in by hand. They also derived the conditions under which such a trail can form and established a connection between the trail deceleration and the cooling luminosity. Together, these studies contribute to a theoretical picture of how a runaway SMBH shapes the observed features.

An important feature of RBH-1 is the highly linear morphology of its trail. This may place constraints on turbulence in the diffuse CGM through which it passes. Although CGM kinematics can be probed through absorption or diffuse emission lines, such measurements have historically been difficult: the former rely on the availability of quasar sightlines, while the latter are often very faint and require very deep integral-field unit (IFU) observations. The trails of runaway black holes therefore open up a possible new approach to probing turbulence in the diffuse CGM, whose kinematics are otherwise difficult to measure.

In this work, we focus on the origin and morphology of the shock front. We use a suite of idealized hydrodynamic simulations of a runaway black hole moving through a uniform medium, exploring different properties of the wind launched by the black hole. By comparing the resulting morphology of the shock front, as well as that of the trail to a limited extent, with the observed features, we constrain the range of wind properties that could reproduce the observations. We also placed an upper bound on the turbulent velocity of the CGM through which RBH-1 passes, based on the linear geometry of its trail. Finally, we use theoretical estimates to relate the required fluxes and total ejected mass to physically plausible wind sources.

We organize the paper as follows. In \sref{S:methods}, we introduce the simulation setup and numerical methods. In \sref{sec:wind}, we present simple scaling relations that estimate the fluxes required to reproduce the shock front. We discuss the variations in wind properties and their effects on the morphology of the shock front in \sref{s:run_variation}, and characterize the most successful models in \sref{s:success}. In \sref{s:dis}, we connect the wind parameters of the successful models to physically plausible mechanisms and describe the limitations of the models. Finally, we summarize our results and discuss the limitations of our models in \sref{s:summary}.

\section{Methodology}
\label{S:methods}
We perform idealized simulations of an initially uniform gas box influenced by a runaway black hole using the \textsc{GIZMO} code\footnote{A public version of this code is available at \href{http://www.tapir.caltech.edu/~phopkins/Site/GIZMO.html}{\textit{http://www.tapir.caltech.edu/$\sim$phopkins/Site/GIZMO.html}}} \citep{Hopkins2015mfm} in its meshless finite mass (MFM) mode. The MFM method combines key advantages of grid-based and smoothed-particle hydrodynamics (SPH) methods within a Lagrangian, mesh-free Godunov framework. The numerical implementation and validation tests are described in detail in methodological papers on hydrodynamics and self-gravity \citep{Hopkins2015mfm}. The simulations include the FIRE-2 model \citep{Hopkins+FIRE2018} for gas cooling over the temperature range $10-10^{10}$ K, with photo-electric and photo-ionization heating, as well as collisional, Compton, fine-structure, recombination, atomic, and molecular cooling. Star formation and FIRE-2 stellar feedback are enabled, but no stars form during the simulations owing to the idealized initial conditions and the relatively short run time.

\subsection{Initial conditions}\label{sec:ic}
The initial conditions consist of a box of static gas, 200 kpc on each side, with uniform density and temperature. A black hole of mass $2\times10^7 M_\odot$ is placed at the center of the box and assigned an initial velocity. The gas mass resolution is $870\,M_\odot$, while the spawned outflow particles (described in \sref{sec:outflow}) have a mass resolution of $100\,M_\odot$. For the runs with turbulence, the initial conditions are first stirred by turbulence as described in \sref{sec:turb}. A detailed summary of the initial conditions and runs is given in \tref{tab:run}.

\subsection{Outflows from black holes}\label{sec:outflow}
We model winds and jets following the prescription described in \cite{2021MNRAS.507..175S,Su+2023,Weinberger2023,Su+JetHaloMass2024,Su+2025+box}. In this implementation, wind launching is handled through a particle-spawning method: new gas cells, or resolution elements, are created to model the wind material. These spawned cells are assigned specified masses, temperatures, and velocities, which together set the specific energy of the wind. This method provides more direct control over the wind properties and reduces the sensitivity of the injection procedure to the conditions of the neighboring gas identified by the neighbor-finding algorithm\footnote{In the more conventional approach, energy and momentum are coupled to gas particles selected through a neighbor search around the black hole, so the resulting feedback depends on the local gas properties and their spatial distribution. See \cite{Wellons2023} for a comparison of different implementations.}. The same approach also makes it possible to impose enhanced resolution within the wind material itself, which is important for resolving light wind.

The masses of the spawned gas particles used in each run are listed in \tref{tab:run}. Once created, these particles are not allowed to merge back into more massive gas elements until their speeds have fallen below 10\% of the original launch velocity. Each launching event produces a pair of particles simultaneously, with one sampled from the upper hemisphere and the other from the lower hemisphere. Such an event occurs whenever the accumulated wind mass flux reaches twice the target mass of a spawned particle. The particles are initially placed at random positions inside a sphere of radius $r_0$, where $r_0$ is set to the smaller of 10 pc and one half of the distance from the black hole to the nearest gas particle.

For a spawned particle with initial position $(r_0,\theta_0,\phi_0)$ in a jet model with opening angle $\theta_{\rm op}$, we remap the initial velocity direction to $2\theta_{\rm op}\theta_0/\pi$ when $\theta_0<\pi/2$, and to $\pi-2\theta_{\rm op}(\pi-\theta_0)/\pi$ when $\theta_0>\pi/2$. This angular mapping prevents the trajectories of different particles from crossing and therefore preserves a coherent wind geometry. In the runs considered here, most outflows are launched nearly isotropically with $\theta_{\rm op}=90^o$. The run with a fixed black hole and collimated jets has an opening angle of $\theta_{\rm op}=1^\circ$.

The injected winds are specified by prescribed mass, momentum, and energy fluxes. The energy flux is
\begin{align}
\dot{E}_{\rm wind}=\dot{M}_{\rm wind}  \left(\frac{1}{2} V_{\rm wind}^2+ \frac{3kT}{2\mu m_p}  \right).
\end{align}
Here, $V_{\rm wind}$ is the adopted wind velocity, and $\mu$ is the average particle mass in units of the proton mass, $m_p$.

\subsection{Turbulence}\label{sec:turb}
We model turbulent stirring by directly driving turbulence following the turbulent box approach of \citet{2012MNRAS.423.2558B}. The turbulent forcing is generated in Fourier space as an Ornstein--Uhlenbeck process \citep[see][]{2009A&A...494..127S,2010A&A...512A..81F,2010MNRAS.406.1659P}, with the maximum driving wavelength, $\lambda=2\pi/k$, set to $10$ kpc. 
We impose purely solenoidal driving by removing the compressive component of the acceleration field through a Helmholtz decomposition in Fourier space. This choice allows us to isolate the kinematic effects of turbulent eddies on the geometry of the trail while excluding direct compressive forcing, which produces larger gas-density fluctuations and therefore more strongly affects radiative cooling. Moreover, simulations of multiphase halo turbulence find that the evolved velocity field is dominated by solenoidal modes even when the imposed large-scale forcing is compressive \citep{2022MNRAS.514.3139M}.  The resulting field is then transformed back to real space and applied continuously to each resolution element as an acceleration. The resulting time-averaged median 1D Mach numbers are roughly 0.18 and 0.39 for the TurbLow and TurbHi runs, respectively. For the run with turbulent driving, we first drive turbulence for 10 Myr in the initially static gas with the uniform density and temperature described in \sref{sec:ic}, before initiating the runaway black-hole evolution.

\begin{table*}
\begin{center}
 \caption{Physics variations of all simulations}
 \label{tab:run}
\resizebox{17.7cm}{!}
{\movetableright=-1.6in
\begin{tabular}{c|cccc|cc|cccc|ccc}
\hline
\hline
&\multicolumn{4}{c|}{Numerical details} & \multicolumn{2}{c|}{BH}&\multicolumn{4}{c|}{Wind parameters} & \multicolumn{3}{c}{Background gas} \\
\hline
Model          & $\Delta t$ & Box size & $m_{\rm g}$ & $m_{\rm wind}$ & $M_{\rm BH}$ &  $V_{\rm BH}$&  $\dot{M}_{\rm wind}$ & $V_{\rm wind}$ & $T_{\rm wind}$ & opening & $n$ &  $T$ & Turbulence \\
           &Myr          & kpc       & $M_\odot$ & $M_\odot$    & $M_\odot$ & km s$^{-1}$  & $M_\odot {\rm yr}^{-1}$  & km s$^{-1}$ &K &     & cm$^{-3}$  & K  \\ 
\hline
NoWind & 100 &200 & 870 & 100 &2e7 &1000 &0 &N/A& N/A & N/A &1e-3 & 1e6& no\\
\hline
\multicolumn{5}{l}{\textbf{Fixed Momentum Flux, Different Velocity and Mass Flux}} \\
\hline
P2e33-v1000 & 100 &200 & 870 & 100 &2e7 &1000 &0.26 &1000& 1e7 & iso &1e-3 & 1e6& no\\
P2e33-v3000 & 100 &200 & 870 & 100 &2e7 &1000 &0.09 &3000& 1e7 & iso &1e-3 & 1e6& no\\
P2e33-v10000 & 100 &200 & 870 & 100 &2e7 &1000 &0.026 &10000& 1e7 & iso &1e-3 & 1e6& no\\
P2e33-v30000 & 100 &200 & 870 & 100 &2e7 &1000 &0.009 &30000& 1e7 & iso &1e-3 & 1e6& no\\
\hline
\multicolumn{5}{l}{\textbf{Fixed Velocity Different Mass Flux}} \\
\hline
P2e32-v1000 & 100 &200 & 870 & 100 &2e7 &1000 &0.026 &1000& 1e7 & iso &1e-3 & 1e6& no\\
P6e32-v3000 & 100 &200 & 870 & 100 &2e7 &1000 &0.03 &3000& 1e7 & iso &1e-3 & 1e6& no\\
P2e32-v3000 & 100 &200 & 870 & 100 &2e7 &1000 &0.009 &3000& 1e7 & iso &1e-3 & 1e6& no\\
P2e32-v10000 & 100 &200 & 870 & 100 &2e7 &1000 &0.0026 &10000& 1e7 & iso &1e-3 & 1e6& no\\
\hline
\multicolumn{5}{l}{\textbf{Different Geometry}} \\
\hline
P2e33-v3000-Jet-fixBH & 100 &200 & 870 & 100 &2e7 &0 &0.09 &3000& 1e7 & Jet &1e-3 & 1e6& no\\
\hline
\multicolumn{5}{l}{\textbf{Time-Varying Momentum Fluxes}} \\
\hline
\multirow{2}{*}{P2e33-v3000-20/70}
& \multirow{2}{*}{100}
& \multirow{2}{*}{200}
& \multirow{2}{*}{870}
& \multirow{2}{*}{100}
& \multirow{2}{*}{2e7}
& \multirow{2}{*}{1000}
& 0.09
& \multirow{2}{*}{3000}
& \multirow{2}{*}{1e7}
& \multirow{2}{*}{iso}
& \multirow{2}{*}{1e-3}
& \multirow{2}{*}{1e6}
& \multirow{2}{*}{no}\\
&&&&&&&$({\rm only}\, 20$--$70\,\mathrm{Myr})$&&&&&&\\
\multirow{2}{*}{P2e32-v3000-linear}
& \multirow{2}{*}{100}
& \multirow{2}{*}{200}
& \multirow{2}{*}{870}
& \multirow{2}{*}{100}
& \multirow{2}{*}{2e7}
& \multirow{2}{*}{1000}
& 0.009$\times$
& \multirow{2}{*}{3000}
& \multirow{2}{*}{1e7}
& \multirow{2}{*}{iso}
& \multirow{2}{*}{1e-3}
& \multirow{2}{*}{1e6}
& \multirow{2}{*}{no}\\
&&&&&&&$(t/63{\rm Myr})$&&&&&&\\
\multirow{2}{*}{P6e32-v3000-duty}
& \multirow{2}{*}{100}
& \multirow{2}{*}{200}
& \multirow{2}{*}{870}
& \multirow{2}{*}{100}
& \multirow{2}{*}{2e7}
& \multirow{2}{*}{1000}
& 0.03 (9-10th Myr
& \multirow{2}{*}{3000}
& \multirow{2}{*}{1e7}
& \multirow{2}{*}{iso}
& \multirow{2}{*}{1e-3}
& \multirow{2}{*}{1e6}
& \multirow{2}{*}{no}\\
&&&&&&&in every 10 Myr)&&&&&&\\
\multirow{2}{*}{P6e32-v3000-end}
& \multirow{2}{*}{100}
& \multirow{2}{*}{200}
& \multirow{2}{*}{870}
& \multirow{2}{*}{100}
& \multirow{2}{*}{2e7}
& \multirow{2}{*}{1000}
& 0.003 ($<$57 Myr)
& \multirow{2}{*}{3000}
& \multirow{2}{*}{1e7}
& \multirow{2}{*}{iso}
& \multirow{2}{*}{1e-3}
& \multirow{2}{*}{1e6}
& \multirow{2}{*}{no}\\
&&&&&&&0.03 ($>$57 Myr)&&&&&&\\
\multirow{2}{*}{P6e31-v3000-n1e-4-end}
& \multirow{2}{*}{100}
& \multirow{2}{*}{200}
& \multirow{2}{*}{870}
& \multirow{2}{*}{100}
& \multirow{2}{*}{2e7}
& \multirow{2}{*}{1000}
& 0.0003 ($<$57 Myr)
& \multirow{2}{*}{3000}
& \multirow{2}{*}{1e7}
& \multirow{2}{*}{iso}
& \multirow{2}{*}{1e-4}
& \multirow{2}{*}{1e6}
& \multirow{2}{*}{no}\\
&&&&&&&0.003 ($>$57 Myr)&&&&&&\\
\hline
\multicolumn{5}{l}{\textbf{Turbulence}} \\
\hline
\multirow{2}{*}{P6e31-v3000-n1e-4-end-Turblow}
& \multirow{2}{*}{100}
& \multirow{2}{*}{200}
& \multirow{2}{*}{870}
& \multirow{2}{*}{100}
& \multirow{2}{*}{2e7}
& \multirow{2}{*}{1000}
& 0.0003 ($<$57 Myr)
& \multirow{2}{*}{3000}
& \multirow{2}{*}{1e7}
& \multirow{2}{*}{iso}
& \multirow{2}{*}{1e-4}
& \multirow{2}{*}{1e6}
& \multirow{2}{*}{Mdn$(\mathcal{M}_{\rm 1D})\simeq0.18$}\\
&&&&&&&0.003 ($>$57 Myr)&&&&&&\\
\multirow{2}{*}{P6e31-v3000-n1e-4-end-TurbHi}
& \multirow{2}{*}{100}
& \multirow{2}{*}{200}
& \multirow{2}{*}{870}
& \multirow{2}{*}{100}
& \multirow{2}{*}{2e7}
& \multirow{2}{*}{1000}
& 0.0003 ($<$57 Myr)
& \multirow{2}{*}{3000}
& \multirow{2}{*}{1e7}
& \multirow{2}{*}{iso}
& \multirow{2}{*}{1e-4}
& \multirow{2}{*}{1e6}
& \multirow{2}{*}{Mdn$(\mathcal{M}_{\rm 1D})\simeq0.39$}\\
&&&&&&&0.003 ($>$57 Myr)&&&&&&\\
\hline 
\hline
\end{tabular}
}
\end{center}
\begin{flushleft}
A list of simulations studied here with different wind and background gas parameters. The columns list: 
(1) Model name: The name of each model starts with the momentum flux in cgs followed by the velocity in km s$^{-1}$
(2) $\Delta t$: Simulation duration.
(3) Box size of the simulation.
(4) $m_{\rm g}$: The ambient gas mass resolution.
(5) $m_{\rm wind}$: The mass resolution of the spawned wind particles.
(6) $M_{\rm BH}$: Black hole mass
(7) $V_{\rm BH}$: Black hole velocity relative to the ambient medium.
(8) $\dot{M}_{\rm wind}$: Wind mass flux.
(9) $V_{\rm wind}$: Wind velocity.
(10) $T_{\rm wind}$: Wind temperature.
(11) Opening angle: Isotropic or jet.
(12) $n$: Background gas density.
(13) $T$: Background gas temperature.
(14) Turbulence: whether turbulent stirring is enabled and, if so, the average Mach number.
\end{flushleft}
\end{table*}

\section{Analytic expectations for wind-driven bow shocks}
\label{sec:wind}

\begin{figure*}
    \centering
    \includegraphics[width=17cm]{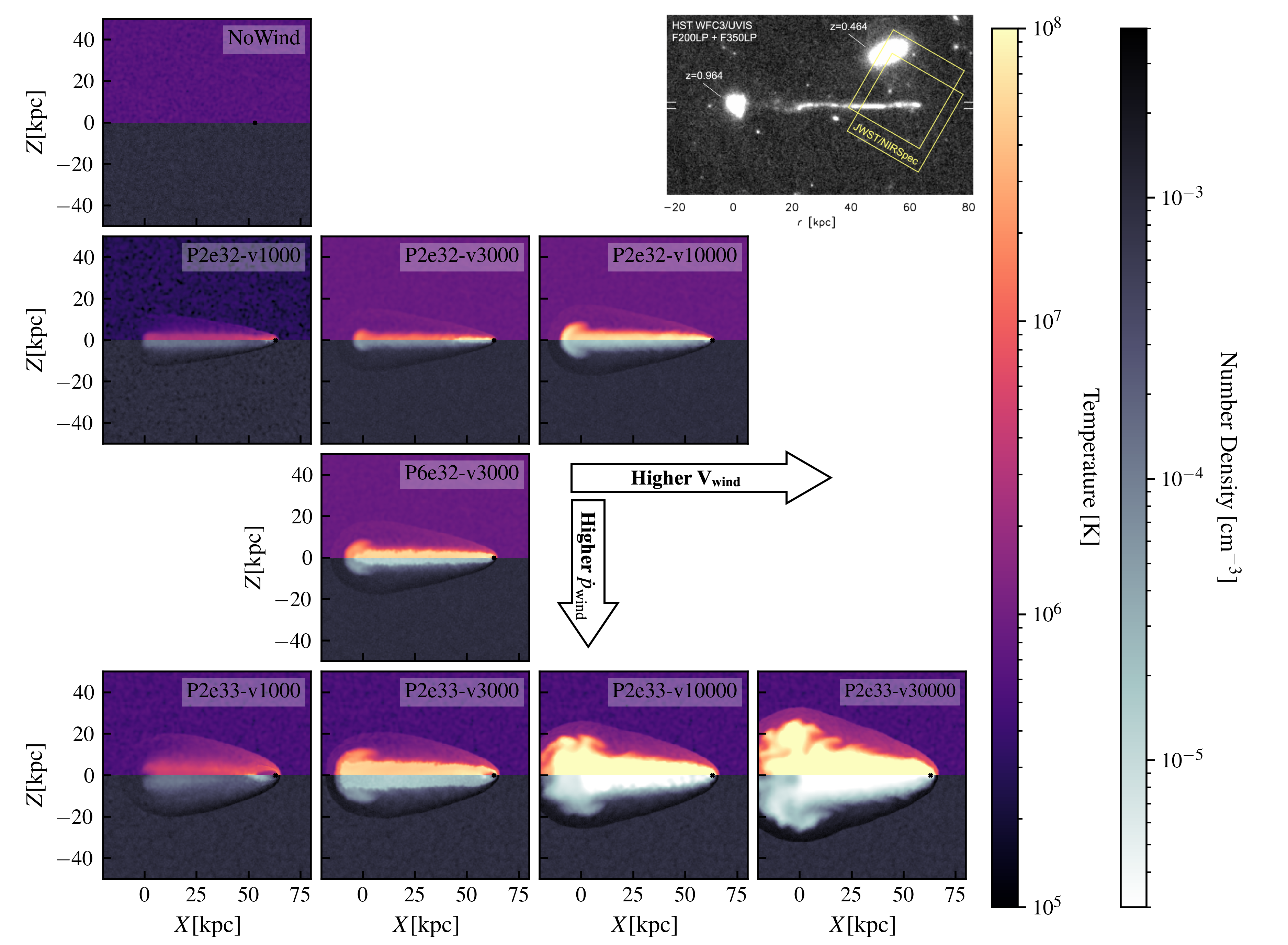}
    \caption{Zoomed-out views of the shock front and trail, shown for comparison with Fig.~1 of \cite{2026ApJ...998L..27V} (reproduced in the upper right). The upper half in each panel
 shows the temperature, while the lower half in each panel shows the number density. The wind velocity increases from left to right (1000-30000 km s$^{-1}$), while the momentum flux increases from top to bottom ($2\times10^{32}-2\times10^{33}$ dyn). The top left corner shows the no-wind case, indicating that an ongoing wind is absolutely necessary to reproduce the bow shock.
 At fixed momentum flux, a higher wind velocity produces a hotter, lower-density bubble. At fixed wind velocity, a higher momentum flux produces a more extended shock front. If the wind velocity exceeds the black hole velocity, a second shock front forms at the opposite end of the trail.} 
    \label{fig:all_zoom_out}
\end{figure*}

\begin{figure*}
    \centering
    \includegraphics[width=17cm]{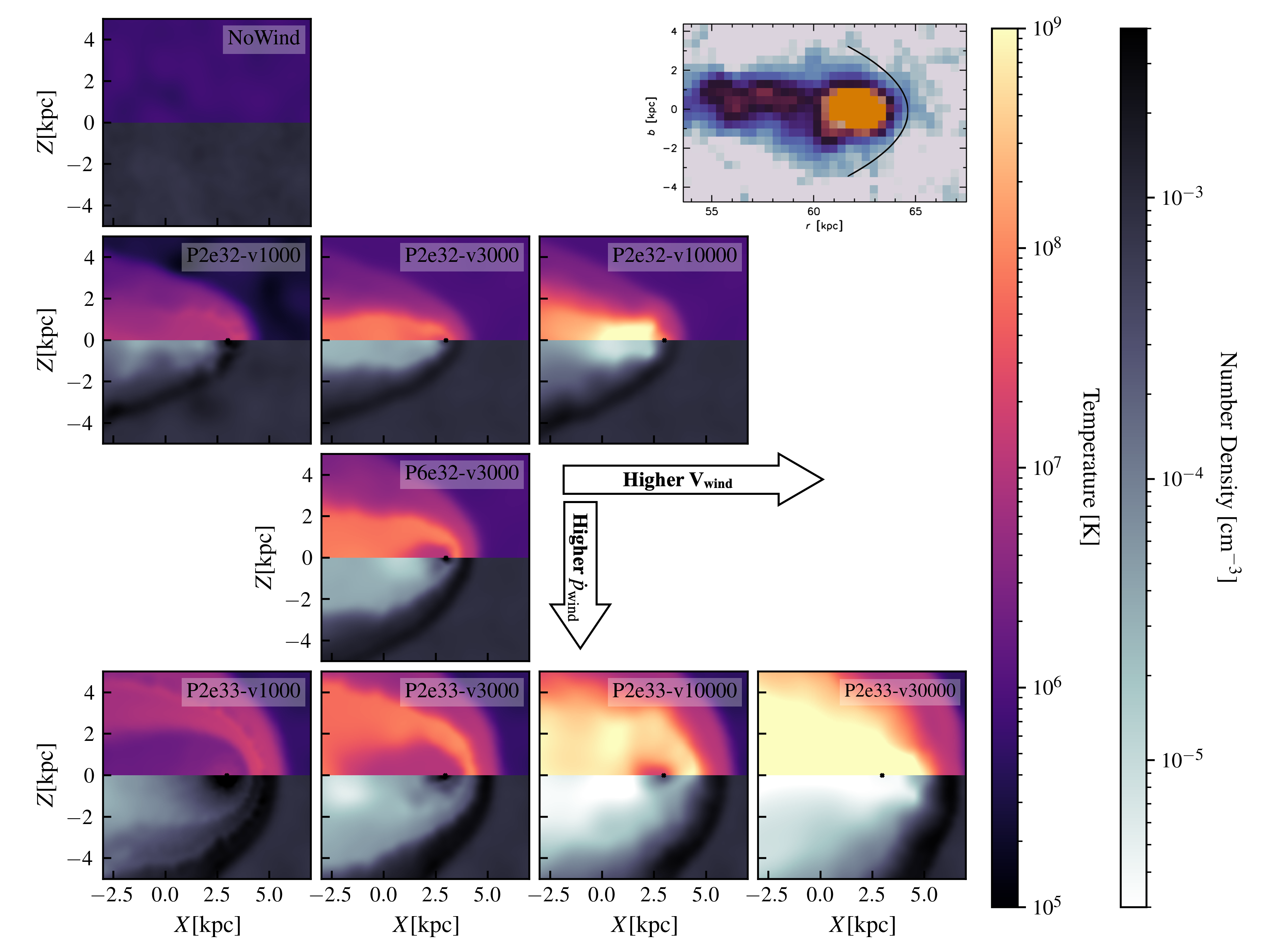}
    \caption{The same as \fref{fig:all_zoom_out}, but zoomed in for comparison with Fig.~3 of \cite{2026ApJ...998L..27V} (reproduced in the upper right). The upper half in each panel
 shows the temperature, while the lower half in each panel shows the number density. The wind velocity increases from left to right (1000-30000 km s$^{-1}$), while the momentum flux increases from top to bottom ($2\times10^{32}-2\times10^{33}$ dyn). Without active wind (upper left), no bow shock will be generated.  At fixed momentum flux, a higher wind velocity produces a hotter, lower-density bubble. At fixed wind velocity, a higher momentum flux produces a more extended shock front. }
\label{fig:all_zoom_in}    
\end{figure*}

\begin{figure*}
    \centering
    \includegraphics[width=17cm]{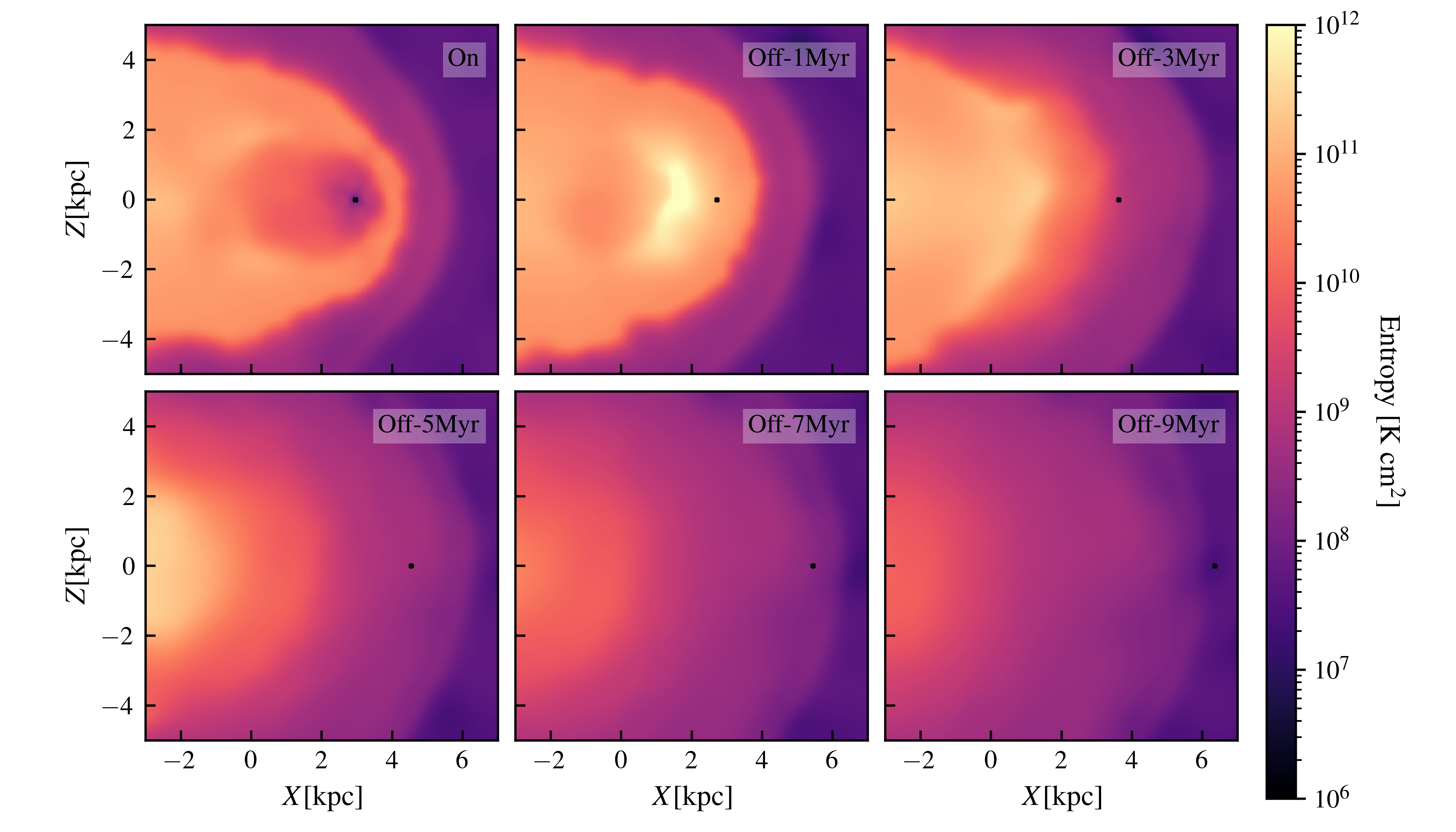}
    \caption{The entropy morphologies of a run with continuous $2\times10^{33}$ dyn momentum flux from a 3000 km s$^{-1}$ wind (P2e33-v3000--20/70) are shown while the wind is active and at different times after it is turned off. When the wind is active, the simulation roughly reproduces the shape of the shock front observed in RBH-1. Three Myr after the wind is turned off, the shock front becomes much more diffuse. By 7--9 Myr after shutdown, the black hole has overtaken the shock front because it moves faster. This shows that an active wind is necessary to sustain a bow shock.
}
    \label{fig:on_off}
\end{figure*}

\fref{fig:all_zoom_out} and \fref{fig:all_zoom_in}  show the zoomed-out and zoomed-in views, respectively, of the no-wind run and the runs with steady isotropic winds. The upper panels show temperature slices, while the lower panels show number-density slices. 

As the wind propagates outward from the central black hole, it inflates a bubble around it. Most of the ambient gas within the bubble is swept into a thin, dense shell whose outer edge defines the shock front. The interior of the bubble is filled with pressurized, reverse-shocked wind material, which has a higher temperature and lower density than the ambient medium.

In the presence of ram pressure, the upstream side of the bubble expands until the cocoon pressure becomes approximately equal to the ram pressure. By contrast, the lateral extent can be larger because it is not directly confined by the headwind. The downstream trail can be the most extended structure, potentially spanning the entire trajectory of the black hole.

\subsection{Momentum-driven}
In the limit where the shock is momentum driven, the momentum flux carried by an isotropic wind is

\begin{equation}
    \frac{d\dot{p}_{\rm wind}}{dA}
    =
    \frac{\dot{p}_{\rm wind}}{4\pi R^2},
\end{equation}

where $\dot{p}_{\rm wind}$ is the total injected momentum rate.

The upstream expansion is halted when the wind momentum flux balances the ram pressure of the ambient medium,

\begin{equation}
    \frac{\dot{p}_{\rm wind}}{4\pi R_{\rm ram}^2}
    =
    \rho v_{\rm ram}^2,
\end{equation}

where $\rho$ is the ambient density and $v_{\rm ram}$ is the relative velocity between the source and the ambient medium.

Therefore, the corresponding stand-off (or ram-pressure) radius is

\begin{equation}
    R_{\rm ram}
    =
    \left(
    \frac{\dot{p}_{\rm wind}}
    {4\pi\rho v_{\rm ram}^2}
    \right)^{1/2},
\label{eq:r_mom}    
\end{equation}

or equivalently,

\begin{align}
    \dot{p}_{\rm wind}
    &=
    4\pi R_{\rm ram}^2
    \rho
    v_{\rm ram}^2\notag\\
    &\simeq
    1.8\times10^{33}
    \left(\frac{\mu}{0.62}\right)
    \left(\frac{n}{10^{-3}\ {\rm cm^{-3}}}\right)
    \left(\frac{R_{\rm ram}}{1.2\ {\rm kpc}}\right)^2\notag\\
    &\left(\frac{v_{\rm ram}}{1000\ {\rm km\,s^{-1}}}\right)^2
    \ {\rm dyn}.
\end{align}

\subsection{Energy-driven}
Assuming an adiabatic, energy-driven bubble expanding into a uniform ambient
medium of density $\rho$, the thin-shell momentum and interior-energy
equations are

\begin{equation}
    \frac{d}{dt}
    \left(
        \frac{4\pi}{3}\rho R^3 \dot{R}
    \right)
    =
    4\pi R^2 P,
\end{equation}

and

\begin{equation}
    \frac{d}{dt}
    \left[
        \frac{4\pi R^3 P}{3(\gamma-1)}
    \right]
    =
    \dot{E}_{\rm in}
    -
    4\pi R^2 P\dot{R},
\end{equation}

where $\dot{E}_{\rm in}$ is the constant mechanical energy-injection rate.

For a monatomic ideal gas, $\gamma=5/3$, these equations give the Weaver solution \citep{1977ApJ...218..377W}

\begin{equation}
    R(t)
    =
    \beta
    \left(
        \frac{\dot{E}_{\rm in}}{\rho}
    \right)^{1/5}
    t^{3/5},
\end{equation}

where

\begin{equation}
    \beta
    =
    \left(
        \frac{125}{154\pi}
    \right)^{1/5}
    \simeq 0.763.
\end{equation}

The corresponding interior pressure is

\begin{align}
    P(t)
    &=
     \frac{7}{25}
    \beta^2
    \dot{E}_{\rm in}^{2/5}
    \rho^{3/5}
    t^{-4/5} \notag\\
   &=
    \frac{7}{25}
    \beta^{10/3}
    \dot{E}_{\rm in}^{2/3}
    \rho^{1/3}
    R^{-4/3}.
\end{align}

The upstream expansion becomes strongly affected by ram pressure when

\begin{equation}
    P(R_{\rm ram})
    =
    \rho v_{\rm ram}^2,
\end{equation}

which gives

\begin{align}
    R_{\rm ram}
    &=
    \beta^{5/2}
    \left(
        \frac{7}{25}
    \right)^{3/4}
    \left(
        \frac{\dot{E}_{\rm in}}
        {\rho v_{\rm ram}^3}
    \right)^{1/2}\notag\\
    &\simeq
    0.196
    \left(
        \frac{\dot{E}_{\rm in}}
        {\rho v_{\rm ram}^3}
    \right)^{1/2}\notag\\
        &\simeq
    0.139
    \left(
        \frac{\dot{p}_{\rm wind}V_{\rm wind}}
        {\rho v_{\rm ram}^3}
    \right)^{1/2}
    .
    \label{eq:r_eng}
\end{align}

Equivalently,

\begin{align}
    \dot{E}_{\rm in}
    &\simeq
    26.1\,
    \rho v_{\rm ram}^3 R_{\rm ram}^2\notag\\
    &\simeq
    3.7\times10^{41}
    \left(\frac{n}{10^{-3}\ {\rm cm^{-3}}}\right)
    \left(\frac{R_{\rm ram}}{1.2\ {\rm kpc}}\right)^2\notag\\
    & \left(\frac{v_{\rm ram}}{1000\ {\rm km\,s^{-1}}}\right)^3
    \ {\rm erg\,s^{-1}}.
\end{align}

\section{Numerical results}
\label{s:run_variation}
Below, we briefly discuss how the cocoon and trail depend on the different wind parameters.
\subsection{No Wind}

In the absence of a black hole wind, the gas is barely affected, indicating that the black hole's purely gravitational influence is negligible on the scales considered \citep[consistent with the analytic estimate in][]{2026ApJ...998L..27V}.
 The Bondi radius is
$r_{\rm Bondi} \sim 4\,{\rm pc}
\left(\frac{T}{10^6\,{\rm K}}\right)^{-1}
\left(\frac{M_{\rm BH}}{2\times10^7\,M_\odot}\right),$ beyond which the gravitational influence of the black hole becomes negligible.  Due to the head wind, the gravitationally affected radius is even smaller at 
\begin{align}
r=\frac{2GM_{\rm BH}}{v^2}\sim 0.17 {\rm \,pc} \left(\frac{v_{\rm BH}}{1000{\rm km s}^{-1}}\right)^{-2}
\label{eq:rgrav}
\end{align}
This radius is much smaller than the observed scale of the shock front.

We also find, however, that the presence of pressurized hot gas alone cannot sustain the shock front for long in the absence of ongoing wind injection from the SMBH. In our simulations, the wind must either remain active or have been turned off less than $\sim 3$ Myr ago. \fref{fig:on_off} shows the entropy profile of the P2e33-v3000 run while the wind is active and at several times after it is turned off. While the wind is active, the simulation produces a bow shock whose shape resembles that observed by \cite{2026ApJ...998L..27V}, enclosing hot, pressurized gas. By 3 Myr after the wind is turned off, the shock front has become substantially more diffuse. By 7 - 9 Myr after shutdown, the black hole has overtaken the decelerating shock front.

\subsection{Identical wind flux but different velocity}
We consider two sets of runs with fixed momentum flux but varying wind velocity. The first set has a momentum flux of $2\times10^{33}\,{\rm g\,cm\,s^{-2}}$ and includes P2e33-v1000, P2e33-v3000, and P2e33-v10000. The second set has a momentum flux of $2\times10^{32}\,{\rm g\,cm\,s^{-2}}$ and includes P2e32-v1000, P2e32-v3000, and P2e32-v10000.

Within each set, runs with different wind velocities produce shock fronts of approximately the same size up to $V_{\rm wind}\sim 10000\,{\rm km\,s^{-1}}$, consistent with \Eqref{eq:r_mom}. At still higher velocities, a slightly more extended shock front is generated. For these high-specific-energy, low-density winds, shock heating occurs at a smaller radius, causing the bubble to become energy-driven. The radius then follows \Eqref{eq:r_eng} and increases with wind velocity.

In all cases, higher wind velocities produce hotter, lower-density wind cocoons. Moreover, when the wind velocity exceeds the black hole velocity of $1000\,{\rm km\,s^{-1}}$, an additional shock front develops behind the black hole, on the side opposite to its direction of motion. The size of this secondary shock front increases with wind velocity.

\subsection{Identical wind velocity with different flux}

At fixed wind velocity, a higher momentum flux produces a larger shock front, as expected from \Eqref{eq:r_mom} and \Eqref{eq:r_eng}. The temperature within the bubble, however, remains approximately unchanged because it is set by the post-shock temperature associated with the wind velocity.

\begin{figure*}
    \centering
    \includegraphics[width=17cm]{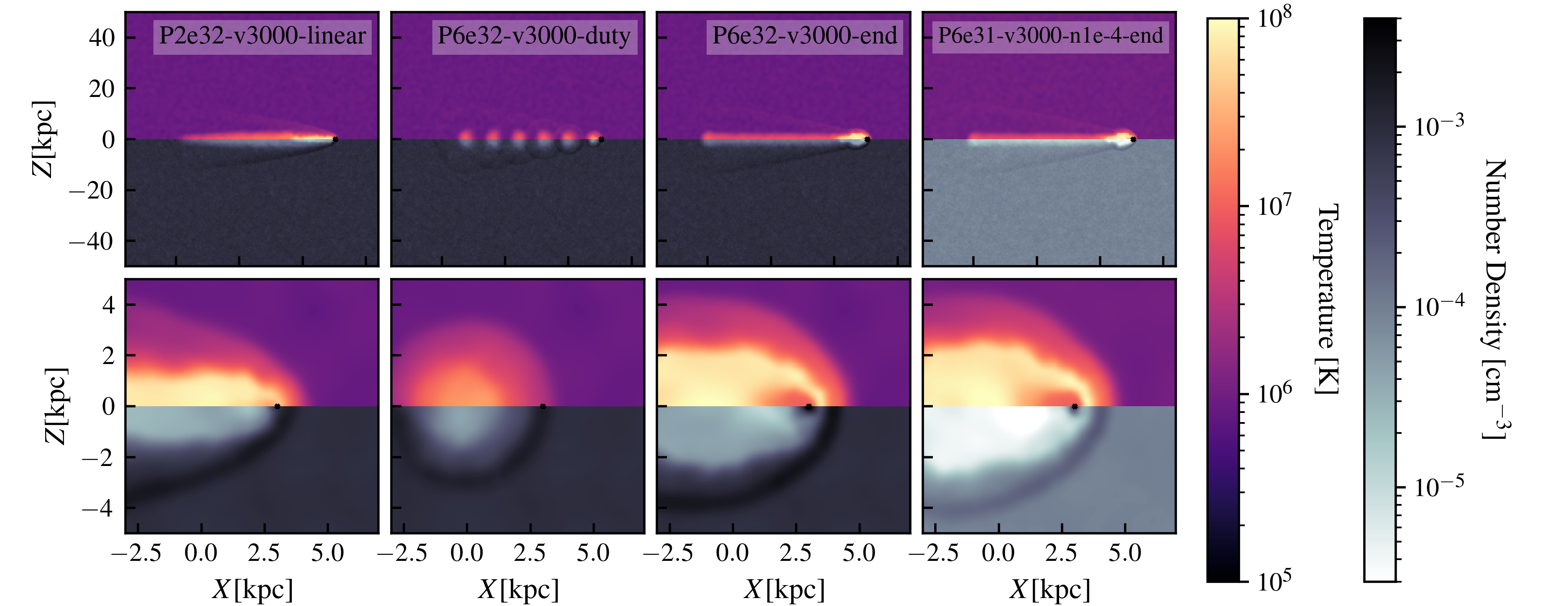}
    \caption{The same as \fref{fig:all_zoom_out} and \fref{fig:all_zoom_in}, but for runs with different time-varying momentum flux (constant wind velocity of 3000 km s$^{-1}$). The upper half in each panel shows the temperature, while the lower half in each panel shows the number density. The width of the trail directly correlates with the wind momentum flux at the time when the black hole passes through each location. The current shape of the shock front depends only weakly on the previous flux history, but mainly on the current momentum flux. Lowering the background density and the momentum flux by the same fraction (3rd and 4th columns) results in similar shock fronts and trails.   }
\label{fig:duty}    
\end{figure*}

We also consider a set of runs exploring different wind-flux histories as demonstrated in \fref{fig:duty}. In P2e32-v3000-linear, the momentum flux increases linearly until it reaches $2\times10^{32}\,{\rm dyn}$. In P6e32-v3000-duty, the wind is turned on with a momentum flux of $6\times10^{32}\,{\rm dyn}$ during the final 1 Myr of every 10 Myr cycle. In P6e32-v3000-end and P6e31-v3000-n1e-4-end, the momentum flux remains at 10\% of its maximum value for the first 57 Myr and then increases to its full value.

Unsurprisingly, the width of the trail is strongly affected by the wind momentum flux at the time when the black hole passes through each location. The final shape of the shock front, however, depends only weakly on the previous flux history. The main exception is the intermittent-flux run with a 10 Myr duty-cycle period, P6e32-v3000-duty, for which the shock front is significantly smaller than that in the run with the flux held constant at its peak value, P6e32-v3000. This occurs because the 1 Myr active phase is too short for the bubble to reach the saturated solution in which the wind momentum flux balances the ram pressure before the wind is turned off again.

We also find that reproducing the observed shock front requires the wind to be currently active. Otherwise, the bubble dissipates rapidly and the shock front becomes blurred.

\subsection{Jet}
\begin{figure*}
    \centering
    \includegraphics[width=13cm]{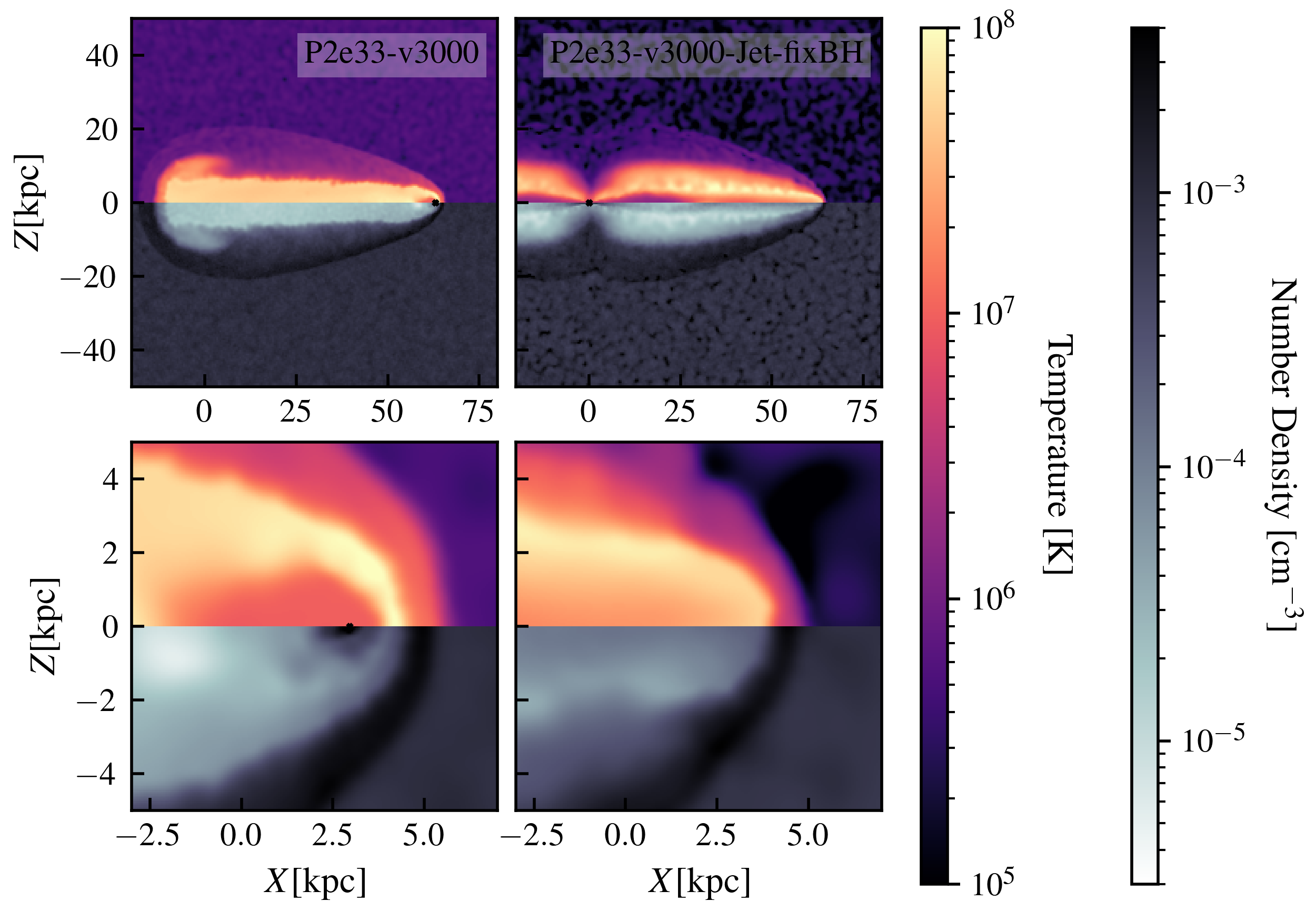}
    \caption{The same plots as in \fref{fig:all_zoom_out} and \fref{fig:all_zoom_in}, but for winds from a moving SMBH with $v_{\rm BH}=1000$ km s$^{-1}$ (left) and a stationary BH with collimated (jet) outflow (right). The two runs have identical momentum fluxes and wind or jet velocities. The upper half in each panel shows the temperature, while the lower half in each panel shows the number density.  A collimated jet with the same momentum flux produces a slightly narrower shock front but a trail that is wider at the base.
}
    \label{fig:jet}
\end{figure*}

\label{s:geometry}
For most of the runs, we adopt isotropic winds from a runaway BH. We also test a static, non-runaway black hole launching a jet. The resulting shock-front geometry, as shown in \fref{fig:jet}, is very similar to that produced by the runaway black hole. It, however, creates a trail that is wider at its base and tapers toward the shock front, a morphology that may be difficult to eliminate by varying the jet flux with time. The cocoon is also visible on both sides.

\subsection{The most successful runs}
\label{s:success}

Among the variations considered, the two runs in which the momentum flux remains at 10\% of its maximum value for the first 57 Myr and then increases to its full value of $6\times10^{32} (n/10^{-3} {\rm cm}^{-3})$ dyn (P6e32-v3000-end and P6e31-v3000-n1e-4-end) most successfully reproduce both the size of the shock front and the extent of the trail seen in previous HST and JWST observations. As shown in the 3rd and 4th columns of \fref{fig:duty}, these runs produce an upstream shock front approximately $2$ kpc in size and a relatively thin trail along the black hole’s trajectory. In both runs, the wind momentum flux remains at 10\% of its maximum value for the first 57 Myr and then increases to its full value. The two runs differ in their ambient gas densities, which are $10^{-3}\,{\rm cm}^{-3}$ and $10^{-4}\,{\rm cm}^{-3}$, respectively. The normalization of the wind momentum flux is scaled linearly with the ambient gas density, in order to keep the standoff radius roughly constant.

\begin{figure*}
    \centering
    \includegraphics[width=15cm]{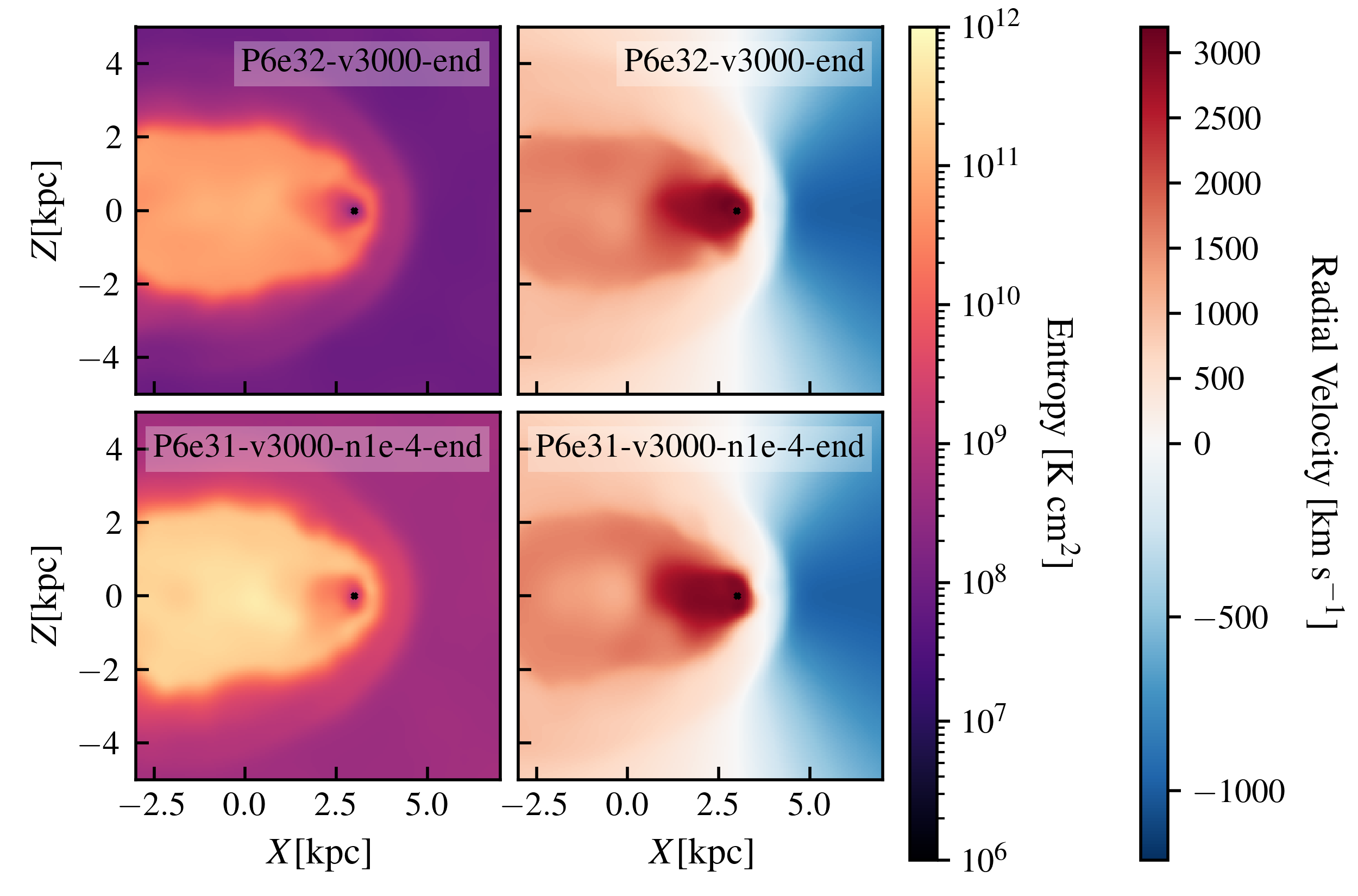}
    \caption{The entropy (left) and radial-velocity (right) morphologies of the most successful runs, P6e32-v3000-end and P6e31-v3000-n1e-4-end. The zoom-out and zoom-in views of the runs can also be seen in the 3rd and 4th columns of \fref{fig:duty}, where the trail width and the size of the shock front resemble the observations. Their background densities are $10^{-3}$ and $10^{-4}\,{\rm cm}^{-3}$, respectively, and their maximum momentum fluxes (during the final $\sim6\,{\rm Myr}$) follow $\dot{p}_{\rm max}=6\times10^{32}\,{\rm dyn}\,(n/10^{-3}\,{\rm cm}^{-3})$. The bubble is filled with high-entropy gas, except within $\sim 0.5$ kpc of the black hole, where the wind ejecta retain their initial velocity and remain denser and cooler before being shock-heated.}
    \label{fig:success}
\end{figure*}

\begin{figure*}
    \centering
    \includegraphics[width=17cm]{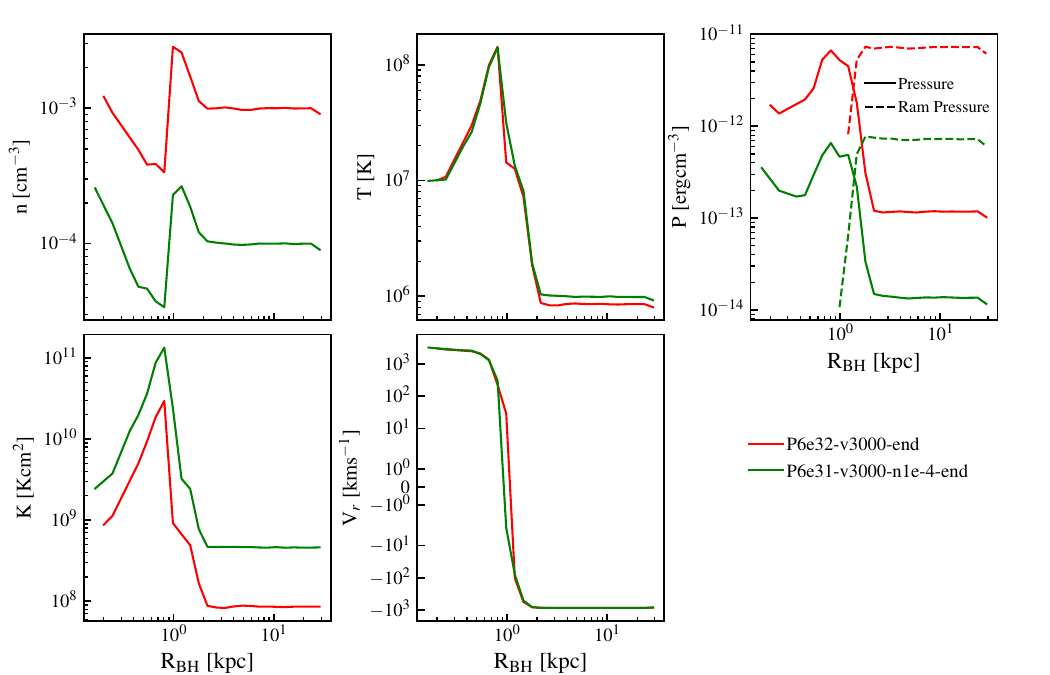}
    \caption{The number-density, temperature, pressure, entropy, and radial-velocity profiles of the two most successful runs shown in \fref{fig:success}, P6e32-v3000-end and P6e31-v3000-n1e-4-end. Their background densities are $10^{-3}$ and $10^{-4}\,{\rm cm}^{-3}$, respectively, and their maximum momentum fluxes during the final $\sim6\,{\rm Myr}$ follow $\dot{p}_{\rm max}=6\times10^{32}\,{\rm dyn}\,(n/10^{-3}\,{\rm cm}^{-3})$.  The profiles are calculated within the upstream region satisfying $\ell/r<1/\sqrt{2}$, corresponding to an opening angle of $45^\circ$, where $\ell$ is the lateral distance to the BH trajectory and $r$ is the distance to the black hole. The bubble is filled with high-entropy gas, except within $\sim 0.5$ kpc of the black hole, where the wind ejecta retain their initial velocity and remain denser and cooler before being shock-heated. As expected, the pressure inside the bubble approximately matches the ram pressure outside the bubble.}
    \label{fig:success_profile}
\end{figure*}

\fref{fig:success} shows the entropy and radial-velocity morphologies of the successful runs. The density, temperature, pressure, entropy, and radial-velocity profiles along the upstream direction are further quantified in \fref{fig:success_profile}. The profiles are calculated within the upstream region satisfying $\ell/r<1/\sqrt{2}$, corresponding to an opening angle of $45^\circ$, where $\ell$ is the lateral distance to the BH trajectory and $r$ is the distance to the black hole.  Consistent with the temperature and density maps and profiles, the bubble is filled with high-entropy gas. It is also worth noting that a $\sim 0.5\,{\rm kpc}$ region around the black hole has slightly lower entropy because of the lower initial temperature and higher density of the wind ejecta.

In the radial-velocity maps, the region extending from $\sim 0.5\,{\rm kpc}$ upstream to $\sim 2\,{\rm kpc}$ downstream is filled with outflowing gas whose radial velocity reflects the initial wind velocity. Along the upstream direction, the gas transitions from the outflowing wind near the black hole to inflowing gas driven by the headwind at larger distances, with the transition occurring within $\sim 2\,{\rm kpc}$ of the black hole. The pressure inside the shock front is also comparable to the ram pressure outside the shock front, consistent with the picture described in \sref{sec:wind}.

\subsection{Constraints on the Turbulence}

\begin{figure*}
    \centering
    \includegraphics[width=17cm]{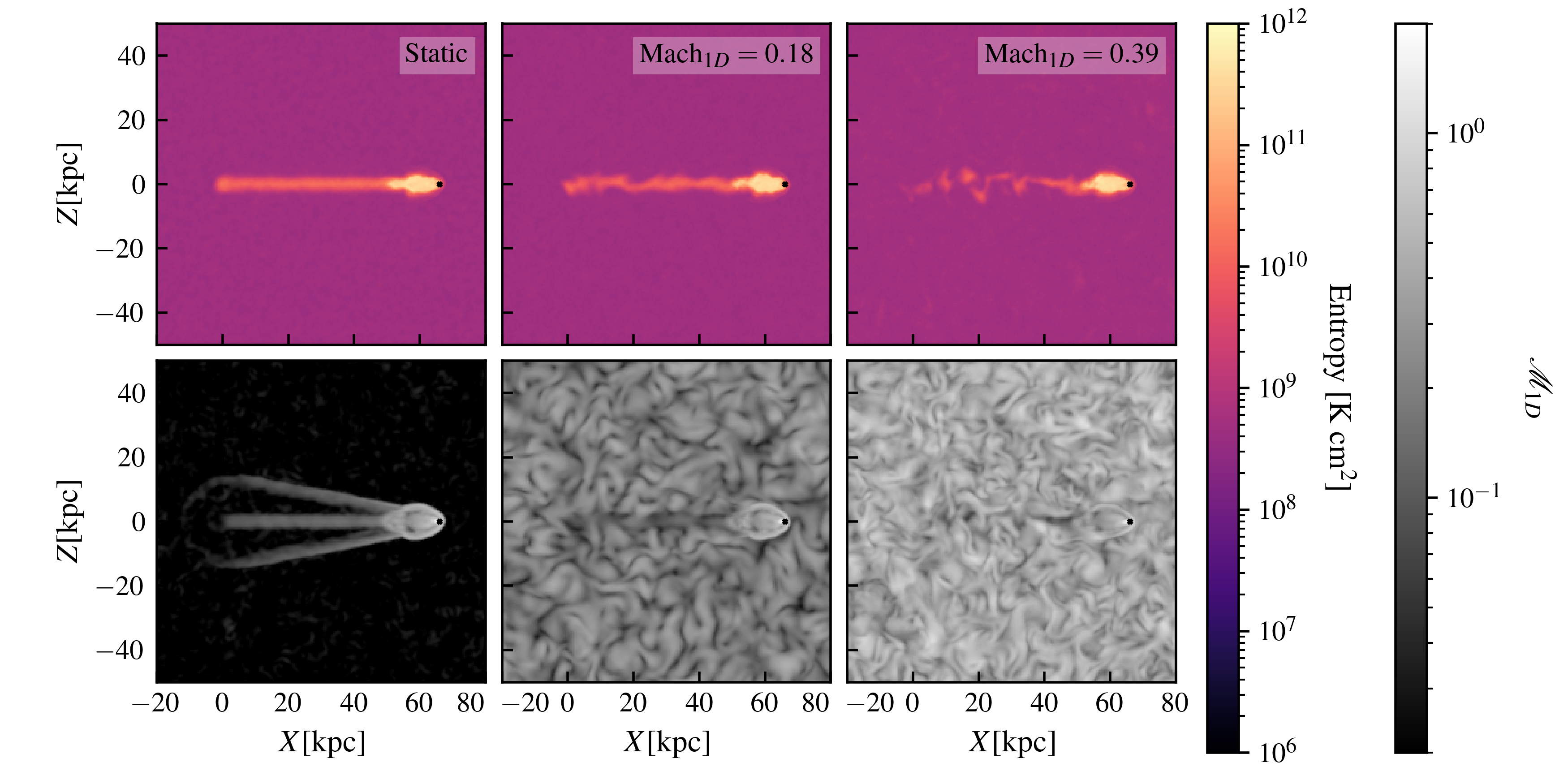}
    \caption{Entropy (upper panels) and Mach number (lower panels) morphologies for runs with different levels of turbulence.
All three runs have the same momentum as the most successful run (`P6e31-v3000-n1e-4-end'), whose momentum flux and wind velocity produce a good match to the observed RBH1 properties (as discussed in \sref{s:success}). The only difference among the three runs is the turbulence of the ambient gas. The Mach number shown in each label denotes the median Mach number in the ambient medium.
   Given the quasi-linear morphology observed in the trail of the runaway SMBH \citep{2023ApJ...946L..50V}, we infer an upper bound on the 1D turbulent velocity of $\sigma_{\rm 1D}\lesssim 30$ km s$^{-1}$ for $10^6$ K gas (a 1D Mach number of $\lesssim 0.18$), above which the trail becomes strongly disturbed.
}
    \label{fig:Turb}
\end{figure*}

While the fiducial runs start with initially static ambient CGM gas, we also perform several variations with different levels of initial turbulence and turbulent driving, as detailed in \tref{tab:run}. The resulting Mach number and entropy distributions are shown in \fref{fig:Turb}. Given the quasi-linear morphology of the runaway SMBH trail \citep{2023ApJ...946L..50V}, we infer an upper limit on the ambient turbulence of a 1D Mach number $\lesssim 0.18$, corresponding to a 1D velocity dispersion $\lesssim 30$ km s$^{-1}$ for a $10^6$ K background. Above this level, the trail becomes strongly disturbed.

\section{Discussion}\label{s:dis}

\subsection{The Origin of the Wind}
\label{s:origin}
From the previous figures, we conclude that an ongoing wind from the black hole is necessary to reproduce the observed morphology. Based on the parameters of the successful runs, we now discuss the possible origins of the wind.

As a recap, both successful runs have a wind velocity of $3000\,{\rm km\,s^{-1}}$. For the run with an ambient gas density of $10^{-3}\,{\rm cm}^{-3}$, the maximum mass, momentum, and energy fluxes  (active at the time the bow shock is observed) are $0.03\,M_\odot\,{\rm yr}^{-1}$, $6\times10^{32}\,{\rm dyn}$, and $10^{41}\,{\rm erg\,s^{-1}}$, respectively. The total mass ejected in the wind is $\sim3.5\times10^5\,M_\odot$ by 63 Myr, when the black hole reaches its observed distance from the galaxy. For the run with an ambient gas density of $10^{-4}\,{\rm cm}^{-3}$, all three fluxes and the total ejected mass are lower by an order of magnitude ( $0.003\,M_\odot\,{\rm yr}^{-1}$, $6\times10^{31}\,{\rm dyn}$, and $10^{40}\,{\rm erg\,s^{-1}}$) so the total ejected mass is $\sim3.5\times10^4\,M_\odot$ by 63 Myr.

\subsubsection{Active galactic nucleus?}
Feedback powered directly by accretion onto the supermassive black hole is a potential source of the wind. Assuming a standard thin accretion disk, the disk mass is roughly
\begin{align}
M^{\rm thin}_d &\sim \frac{H}{R} M_{\rm BH}\notag\\
&\sim 2\times10^5\,M_\odot \left(\frac{M_{\rm BH}}{2\times10^7M_\odot}\right) \left(\frac{H/R}{0.01}\right)
\end{align}
as constrained by the Toomre $Q$ stability argument \cite[e.g.,][]{1981ARA&A..19..137P,1985apa..book.....F}, where $H/R$ is the aspect ratio of the accretion disk, which can range  $\sim0.001-0.1$ \citep[e.g.,][]{2015MNRAS.453L..46K,2008ApJ...687L..25S,
2021MNRAS.507..983L}. 
Assuming it is in the sub-Eddington regime, \cite{2007PhRvL..99d1103L} estimates a very similar disk mass carried by a runaway SMBH using an $\alpha$-disk model:
\begin{align}\label{eq:acc}
&M^{\alpha}_d
\sim 2.3\times10^5\,M_\odot
\left(\frac{\alpha}{0.3}\right)^{-0.8}
\left(\frac{\epsilon_{\rm rad}}{0.1}\right)^{-0.6}
\notag\\
&
\left(\frac{L_{\rm BH}}{0.01L_{\rm Edd}}\right)^{0.6}
\left(\frac{M_{\rm BH}}{2\times10^7\,M_\odot}\right)^{2.2}
\left(\frac{V_{\rm BH}}{1000\,{\rm km\,s}^{-1}}\right)^{-2.8}.
\end{align}
A later, more detailed accretion-disk model accounting for disk self-gravity through a marginally stable $Q=1$ outer region \citep{2008MNRAS.390.1311B} also yields a similar disk mass carried by the runaway SMBH for the same parameters, by integrating the resulting piecewise surface-density profile within the gravitationally bound radius.

Meanwhile, for $M_{\rm BH}=2\times10^7\,M_\odot$, the accretion timescale can be estimated to zeroth order using $L_{\rm BH}=\epsilon_{\rm rad}\dot{M}_{\rm BH}c^2$:
\begin{align}
t_{\rm acc}
=&
\frac{M_d}{\dot{M}_{\rm BH}}
=
\frac{\epsilon_{\rm rad}M_d c^2}{L_{\rm BH}}
\notag\\
\simeq&
4.7\times10^{7}\,{\rm yr}\,\notag\\
&\left(\frac{M_d}{2.3\times10^{5}\,M_\odot}\right)
\left(\frac{L_{\rm BH}}{0.01L_{\rm Edd}}\right)^{-1}
\left(\frac{\epsilon_{\rm rad}}{0.1}\right),
\end{align}
where $\epsilon_{\rm rad}$ is the radiative efficiency. A value of $\epsilon_{\rm rad}\sim0.1$ is a reasonable approximation for black holes radiating at $\sim0.01L_{\rm Edd}$ \citep[e.g.,][]{2012MNRAS.427.1580X}.
The estimated accretion timescale is therefore comparable to the black hole's propagation time of $\sim63$ Myr, suggesting that the carried disk could plausibly sustain accretion over this period.

Note that the depletion timescale estimated above accounts only for accretion onto the black hole. If the wind material is also supplied by the carried accretion disk, wind mass loss must be included in the mass budget. Assuming a constant accretion rate of $0.01\dot{M}_{\rm Edd}$ over 63 Myr, the total accreted mass is $\simeq 3.1\times10^5\,M_\odot$. Combined with the total wind ejecta, the required initial disk mass is $\simeq (3.5\text{--}6.6)\times10^5\,M_\odot$, depending on the ambient density. This remains within a factor of a few of the fiducial disk-mass estimate in \Eqref{eq:acc}, given the uncertainties in the disk parameters. In reality, some of the wind mass on the scales we resolve may be swept up from ambient gas rather than directly come from the accretion disk.

Assuming an Eddington ratio of 0.01, the radiative luminosity is $\sim2.8\times10^{43}\,{\rm erg\,s}^{-1}$. The required peak wind powers are $\sim10^{40}$--$10^{41}\,{\rm erg\,s}^{-1}$, so a mechanical output equivalent to only a subpercent fraction of the radiative luminosity would be sufficient.

Indeed, mechanical efficiencies up to of order unity can be achieved in low-Eddington accretion flows around rapidly spinning black holes in the magnetically arrested state \citep{2011MNRAS.418L..79T}, exceeding the fiducial radiative efficiency adopted here. For a fixed wind power, such higher mechanical efficiencies would reduce the required black hole accretion rate and ease the disk mass-budget constraint.

We caution, however, that the lack of an X-ray counterpart noted by \citet{2023ApJ...946L..50V} led them to tentatively conclude that the runaway SMBH is not currently radiatively active. Nevertheless, subsequent HST imaging reveals a compact UV source at the expected SMBH position, providing tentative evidence for accretion activity \citep{2026ApJ...998L..27V}. The association of this source with the SMBH remains to be confirmed.

We note that the expected X-ray luminosity is only $L_{2-10\,{\rm keV}}\sim2\times10^{42}\,{\rm erg\,s}^{-1}$ at 1\% Eddington ($L_{\rm bol}\sim2.8\times10^{43}\,{\rm erg\,s}^{-1}$), assuming that $\lesssim7\%$ of the bolometric luminosity emerges in the rest-frame 2 to 10 keV band \citep{2007MNRAS.381.1235V,2020A&A...636A..73D}.  Scaling the count estimate of \citet{2023ApJ...946L..50V}, we expect $<3$ source counts in the existing \textit{Chandra} observation, so such a source could possibly have escaped detection. A UV upper limit is harder to derive because the HST difference image is difficult to calibrate \citep{2026ApJ...998L..27V}. 


\subsubsection{Nuclear star cluster?}
We also consider the possibility that the wind is powered by supernovae from a nuclear star cluster carried by the black hole. For a black hole mass of $2\times10^7 M_\odot$, the radius of gravitational influence given by \Eqref{eq:rgrav} is $\sim0.17\,{\rm pc}$. We can estimate the stellar mass associated with the black hole from the mass enclosed within this radius, assuming a characteristic maximum star-cluster surface density. In particular, the nuclear star cluster around Sgr A$^*$ provides a conservative but useful reference point because its structure is also influenced by an SMBH, although its black hole mass, $\sim4\times10^6 M_\odot$, is smaller than that of RBH-1. \cite{2010MNRAS.401L..19H,2019MNRAS.483.5548G} find a characteristic maximum star-cluster surface density of $\Sigma_{\rm star}\sim10^{5.5}\,M_\odot\,{\rm pc}^{-2}$ from empirical compilations of dense stellar systems, including globular clusters, young massive clusters, nuclear star clusters, ultracompact dwarfs, and galaxy spheroids. This is also comparable to the maximum surface density quoted in these studies for the Milky Way nuclear disk, based on the total mass compiled by \cite{2009ApJ...690.1463L} and the profile $\Sigma_{\rm star}\propto(R^2+R_{\rm core}^2)^{-1}$ fitted by \cite{2006ApJ...643.1011P}. This surface density gives $M_{\rm star}^{<0.17{\rm pc}}\sim\Sigma_{\rm star}\pi r^2\sim3\times10^4 M_\odot$.

Other studies suggest a higher surface density for the Sgr A$^*$ nuclear star cluster. Based on the mass profile quoted by \cite{2012ApJ...744...24Y},
\begin{align}
M_{\rm star}(<r)\sim3.2\times10^4 M_\odot
\left(\frac{r}{0.1{\rm pc}}\right)^{2.1},
\end{align}
the enclosed stellar mass within the same radius, $0.17\,{\rm pc}$, is $M_{\rm star}^{<0.17{\rm pc}}\sim10^5 M_\odot$. \cite{2014CQGra..31x4007S} find an even larger enclosed stellar mass of $M_{\rm star}^{<0.17{\rm pc}}\sim3\times10^5 M_\odot$ within $0.17{\rm pc}$. Taken together, these estimates suggest that the enclosed mass of the Sgr A$^*$ nuclear star cluster within $0.17\,{\rm pc}$ is
\begin{align}
M_{\rm star}^{<0.17{\rm pc}}\sim3\times10^4\, \text{--}\, 3\times10^5 M_\odot.
\end{align}

According to the \textsc{STARBURST99} \citep{1999ApJS..123....3L} predictions plotted by \citet{2018MNRAS.473.3717F}, the maximum momentum flux per unit stellar mass at times $\lesssim 10$ Myr following a burst of star formation, assuming a standard Kroupa IMF \citep{2001MNRAS.322..231K}, can reach
\begin{align}\label{eq:momentum_budget_sne}
\frac{\dot{p}}{M_{\rm star}}\sim10^{27}\,{\rm dyn}\,M_\odot^{-1}.
\end{align}
For a star cluster with a mass of $3\times10^4-3\times10^5\,M_\odot$, the total momentum flux supplied by supernovae is therefore
\begin{align}
\dot{p}\sim3\times10^{31} - 3\times10^{32} \,{\rm dyn},
\end{align}
which is roughly comparable to the required wind momentum flux. 

The required mass flux in the successful runs corresponds to ejecting $\sim10\%$ of the total stellar mass through SNe over the $\sim63$ Myr duration in the lower-density case ($10^{-4}$ cm$^{-3}$), assuming a star cluster mass of $3\times10^{5}M_\odot$.
Following \citet{Hopkins+FIRE2018}, the integrated number of core-collapse supernovae per unit stellar mass within $\sim38$ Myr is approximately $0.01\,M_\odot^{-1}$. The ejecta mass fraction can therefore be estimated as
\begin{align}
f_{\rm eject} &= \frac{N_{\rm SN}}{M_\star} \times m_{\rm ejecta}^{\rm CC} \notag\\
&\sim (0.01\,M_\odot^{-1})(10.5\,M_\odot) \sim 10\%,
\end{align}
where $m_{\rm ejecta}^{\rm CC}=10.5\,M_\odot$ is the mean ejecta mass per core-collapse supernova \citep{Hopkins+FIRE2018}. Therefore, ejecting $\sim10\%$ of the total stellar mass through SNe is within a plausible range.  For the higher-density run ($10^{-3}$ cm$^{-3}$), this integrated ejected mass becomes an order-unity fraction of the stellar mass, which is too high.

However, even if the total integrated ejecta mass fits within the available mass budget, the energy flux from a nuclear star cluster should peak within the first $\sim10$ Myr, when the core-collapse SN rate is highest. At 60-70 Myr, the SN rate is at least 1 dex lower than the rate underlying the estimate in \Eqref{eq:momentum_budget_sne}, even when binary star evolution is included \citep{2017A&A...601A..29Z}. Ongoing star formation around the black hole is also difficult, as it would most likely rely on gas carried by the black hole, which would need to form stars before being blown away by winds. Therefore, the star cluster scenario faces substantial challenges.

\subsubsection{Jet without runaway BH?}
We also performed a run testing the scenario in which the shock and the trail are produced by a jet launched from an active galactic nucleus in the host galaxy, without a runaway black hole. As noted in \sref{s:geometry}, jetted feedback with the same flux generally produces a smaller lateral extent of the shock front because of its highly collimated geometry. However, as shown in \fref{fig:jet}, a jet launched from a stationary black hole may produce a broad cocoon near the host galaxy if it is currently active, while a currently active jet would also be required to maintain the sharp contrast at the shock front. It may therefore be even more difficult to tune the AGN flux to simultaneously reproduce the morphology of both the shock front and the trail.

Moreover, \cite{2023ApJ...946L..50V} noted the lack of X-ray and radio detections along the trail, further disfavoring the interpretation that the feature is a jet launched from the host galaxy.

\subsection{Limitation of this study}
We note that the main goal of this work is to explore the parameter space that can reproduce the shape of the shock front and determine whether the required fluxes are physically plausible. We do not attempt to reproduce the specific emission lines observed with HST and JWST. Our simulations also lack the resolution required to capture the cold phase entrained in the trail \citep{2024MNRAS.527.5503O,2026MNRAS.550g1275K}, which is likely the primary source of the observed emission. We also note that the shock front itself may not induce cooling in hot ambient gas. Modeling the cold phase is left for future work.

We use a combination of analytic estimates and carry out a parameter survey using idealized simulations to model different scenarios and constrain required wind fluxes, but we do not explicitly resolve black hole accretion and outflow processes, nor star clusters. Doing so would require much higher-resolution simulations, and is also left for future work.

Finally, we explored a limited range of ambient-gas properties and focused primarily on an initially static, uniform-density medium. We showed in limited experiments that turbulence in the ambient medium can modify the trail structure and that this effect can be used to constrain CGM turbulence. However, we did not include large-scale spatial gradients in the ambient medium, and also neglected processes such as magnetic fields and cosmic rays. These effects may be important in the CGM \citep[e.g.,][]{2021MNRAS.507..175S,Su+JetHaloMass2024,2026MNRAS.545f2060S} and further affect the thermodynamic properties and morphologies of the shock front and trail. Exploring these effects we have neglected is another important topic for future work.

\section{Summary}\label{s:summary}

In this work, we use a suite of idealized simulations of a runaway black hole moving through a uniform medium to determine the wind properties that could reproduce the shock front and trail properties observed for RBH-1 in \cite{2023ApJ...946L..50V,2026ApJ...998L..27V}. We summarize our main results as follows:
\begin{itemize}
\item Reproducing the bow shock requires an ongoing wind launched by the runaway black hole. Because of the high assumed sound speed in the ambient medium ($T\sim10^6$ K) and the high headwind velocity, the radius within which gas is significantly affected by the black hole’s gravity is very small. The gravitational interaction alone therefore cannot reproduce the observed shock front.
\item If the wind is turned off, the shock front quickly becomes much less sharp. It is possible that the wind from the runaway black hole is time-variable, as may be expected if it is powered by AGN activity. Our experiments suggest that reproducing the observed shock front in the RBH-1 system requires the wind to be either currently active, or have shut off no more than $\sim3$ Myr ago.
\item At fixed momentum flux, a higher wind velocity produces a hotter cocoon. If the wind velocity exceeds the headwind velocity, a second shock front forms on the opposite side of the black hole.

\item At fixed wind velocity, a higher momentum flux produces a wider shock front.

\item With moderate tuning, a spherically symmetric wind with a maximum wind mass flux of $0.03\,M_\odot\,{\rm yr}^{-1}(n/10^{-3}\,{\rm cm}^{-3})$ and a momentum flux of $6\times10^{32}\,{\rm dyn}\,(n/10^{-3}\,{\rm cm}^{-3})$, active at the time of observation, can reproduce the observed shock front, where $n$ is the number density of the ambient gas. The implied total injected mass is $3.5\times10^5\,M_\odot\,(n/10^{-3}\,{\rm cm}^{-3})$. 

\item We investigated whether an accreting SMBH could supply the required fluxes. A $2\times10^7\,M_\odot$ SMBH accreting at $1\%$ of the Eddington rate could supply the required mass and momentum fluxes, and sustain the bow shock even if only a sub-percent-level fraction of its accretion power were converted into mechanical feedback. The total integrated wind mass is also within the possible mass range of an accretion disk carried by a runaway BH like RBH-1. A jet from a static black hole produces a broader cocoon near the host galaxy, making it harder to reproduce the trail and the shock front simultaneously.

\item We also investigated the possibility that the wind could be powered by a nuclear star cluster carried by the runaway SMBH. The wind mass and momentum fluxes required at the time of observation ($>60$ Myr after ejection) are difficult to reconcile quantitatively with a star cluster that has a typical surface density and is small enough to remain gravitationally bound to RBH-1, assuming a standard IMF.

\item From the morphology of RBH-1's linear trail, we place an upper bound on the 1D turbulent velocity dispersion of the ambient medium, $\sigma_{\rm 1D}\lesssim 30\,{\rm km\,s^{-1}}$, corresponding to a 1D Mach number of $\lesssim 0.18$ for a gas temperature of $10^6$ K. At higher turbulent velocities, the simulated trail becomes highly disturbed and inconsistent with the observed morphology.
\end{itemize}

Important directions for future work include simultaneously reproducing the shock front and the entrained cold gas in the trail and generating realistic mock HST and JWST observations. We emphasize that RBH-1 is not only a first-of-its-kind runaway black hole candidate, but also opens up new possibilities for using this system and similar objects discovered in the future to constrain the properties of runaway black holes and the feedback powered by either the black hole itself or a nuclear star cluster it carries. Moreover, their shock fronts and trails offer a new approach to probing the density and turbulence of the diffuse CGM in regimes where few, if any, other probes are available.

\section*{Acknowledgements}
    We thank Pieter van Dokkum, Peng Oh, Greg Bryan, and Phil Hopkins for their valuable insights and helpful discussions.  KYS and CAFG were supported by NSF through grants AST-2108230, AST-2307327 and AST-2606015; by NASA through grants 80NSSC22k0809, 80NSSC22K1124 and 80NSSC24K1224; by STScI through grant JWST-AR-03252.001-A; and by BSF through grant \#2024262.  The simulations are done on Frontera with allocation AST22010, and Bridges-2 with Access allocations TG-PHY220027 \& TG-PHY220047.  This work was performed in part at Aspen Center for Physics, which is supported by National Science Foundation grant PHY-2210452. All ideas in this work were developed by the authors, who also wrote the original text. ChatGPT-6 was used during the final stages to identify typos and grammatical errors, improve sentence clarity, and assist with literature searches.

\section*{Data Availability statement}
A publicly available release of \textsc{GIZMO} can be obtained from
\url{http://www.tapir.caltech.edu/~phopkins/Site/GIZMO.html}.
Owing to their large storage requirements, the complete simulation outputs are not publicly archived. The processed datasets used to construct the figures are available from the corresponding author upon reasonable request for scientific purposes.

\bibliography{bridging_scales}{}
\bibliographystyle{aasjournal}

\end{CJK}
\end{document}